# "Atomic Bremsstrahlung":
# retrospectives, current status and perspectives

M. Ya. Amusia

The Racah Institute of Physics, The Hebrew University, Jerusalem 91904, Israel
and
A. F. Ioffe Physical-Technical Institute, St. Petersburg 194021, Russia
e-mail: Amusia@vms.huji.ac.il

**Abstract**

We describe here the "Atomic bremsstrahlung" (AB) – emission of continuous spectrum electromagnetic radiation, which is generated in collisions of particles that have internal deformable structure that includes positively and negatively charged constituents. The deformation of one of or both colliding partners induces multiple, mainly dipole, time-dependent electrical moments that become a source of radiation. The history of AB invention is presented and it's unusual in comparison to ordinary bremsstrahlung (OB) properties are discussed. As examples, fast electron – atom, non-relativistic and relativistic collisions are considered. Attention is given to ion – atom and atom – atom collisions. Specifics of "elastic" and "inelastic" (i.e. radiation accompanied by destruction of collision partners) AB will be mentioned. Attention will be given to possible manifestation of AB in Nature and in some exotic systems, for instance scattering of electrons upon muonic hydrogen. Some cooperative effects connected to AB will be considered. New classical schemes similar to AB will be presented.

## 1. Introduction

According to Electrodynamics, both classical and quantum, continuous spectrum electromagnetic radiation is generated when a particle with charge $e_p$ and mass $m_p$ moves in a static external field (Akhiezer and Berestetsky, 1965; Heitler, 1954). A tiny atom or a molecule, a solid body or a macroscopic magnet, a star or a galaxy can form this static field. In all these cases the radiation created is called *Bremsstrahlung* (BrS). It is emitted by a charged projectile particle that looses its kinetic energy (perhaps, potential also) under the action of the static field of the target.

The properties of this BrS are well investigated, presented and discussed at length in almost all books on Electrodynamics (see e.g. Berestetskii, Lifshits, and Pitaevskii, 1974). Thus, it is well known that BrS intensity is proportional to $e_p^2$ and inversely proportional to $m_p^2$. These features, particularly the latter one, very often serve even as a definition of BrS as radiation, whose intensity depends upon $m_p^2$ as $m_p^{-2}$ (see e.g. Landau and Lifshits, 1988). There are tables of calculated values of BrS intensities generated in electron-atom collisions (see e.g. Pratt et al., 1977, 1981). This type of radiation we will call here Ordinary Bremsstrahlung, or OB.

However, it was suggested already almost forty years ago that there exist another mechanism of generating of the continuous spectrum electromagnetic radiation. For example, Amusia (1965) assumed that in collisions of two atoms giant common resonances can be excited that can decay by emitting not only one or several electrons but a single continuous spectrum photon as well. It was demonstrated by Buimistrov (1974) that in electron – atom collision continuous spectrum radiation could be emitted not only by the incoming electron, but also by a target hydrogen atom, dipolar polarized by the incoming electron (Buimistrov and Trakhtenberg, 1975). The universality of this mechanism that was called *Atomic Bremsstrahlung* (AB) or *Polarization Radiation* (PR) was demonstrated and emphasized in (Amusia et al., 1976). This mechanism was employed in discussing gas discharge phenomenon by Zon (1977) and resonance photoemission from solids by Wendin and Nuroch (1977)

As a retrospective, it is hard for me to refrain from recollecting that the publication of our paper (Amusia et al., 1976) was delayed due to referee's (a rather well known physicist) "impossible!" comment for about one and a half years. Indeed, as any noticeable innovation, PR went through all usual stages, from almost "impossible" at the beginning to almost "self-evident" thirty years later. The name ""Atomic" Bremsstrahlung" was introduced in (Amusia, 1982) as a response to Pratt's paper (Pratt, 1981) in Comments on Atomic and Molecular Physics called "Electron Bremsstrahlung", where the dynamic response of the target atom electrons in the process of emission of radiation was neglected completely. The name AB met a lot of criticism from those who emphasized that it is not a sort of Bremsstrahlung but a new mechanism of photon emission. This point of view led to the name "Polarization radiation" or PR that will be used in this paper along with AB.

PR was also independently and even a little bit earlier discovered in investigating collisions of almost entirely classical objects, namely ions in plasma. It was demonstrated by Gailitis and Tzitovich (1974) that the main source of electromagnetic radiation in plasma are unbound semi-classical objects "ion + neutralizing electrons", which become polarized in collision processes. Time-dependent dipole moment is induced in them that become a source of electromagnetic radiation. Qualitatively, Fig.1 illustrates the classical picture of AB (PR) radiation.

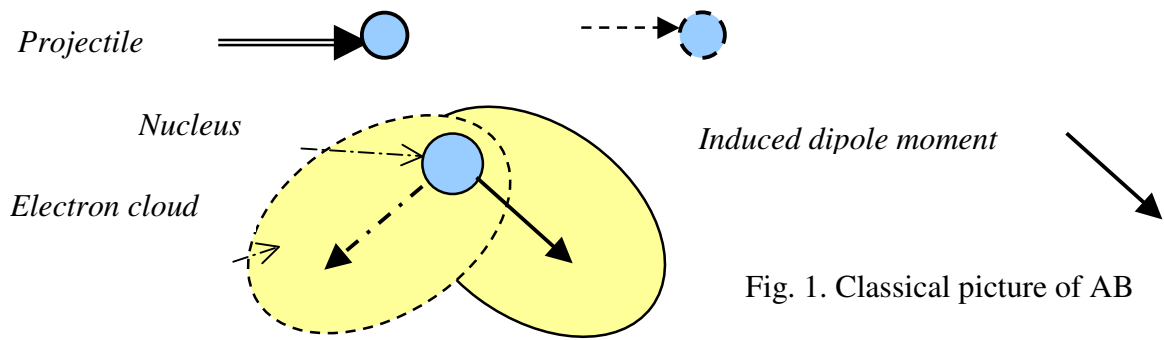

Fig. 1. Classical picture of AB

It depicts a "classical atom" polarized by the incoming charge. In this "classical" atom the positive charge of an ion is surrounded by a number of "classical" electrons. Under the action of the projectile (that is assumed for definiteness to have the same charge as an electron) all target electrons are "pushed out" as is seen in the Figure. So, a dipole (and other multipole) moment are induced. The dipole moment is directed along the axis, on which the "nucleus" and the projectile are located. Due to projectile motion, the induced dipole as well as higher multipole moments are rotating in space and changing their absolute value. As a result, electromagnetic radiation is generated.

Instead of a charge without internal structure, the projectile can be also a "classical" atom or an ion. In this case both, the target and the projectile can emit radiation. Those pictures present mechanisms essential in plasmas. Fig. 2 depicts another "classical" situation: a "classical" atom moves in an external, for definiteness, electrical, static field. Acting differently upon electrons and the nucleus, the external field induces dipole moment in the incoming atom.

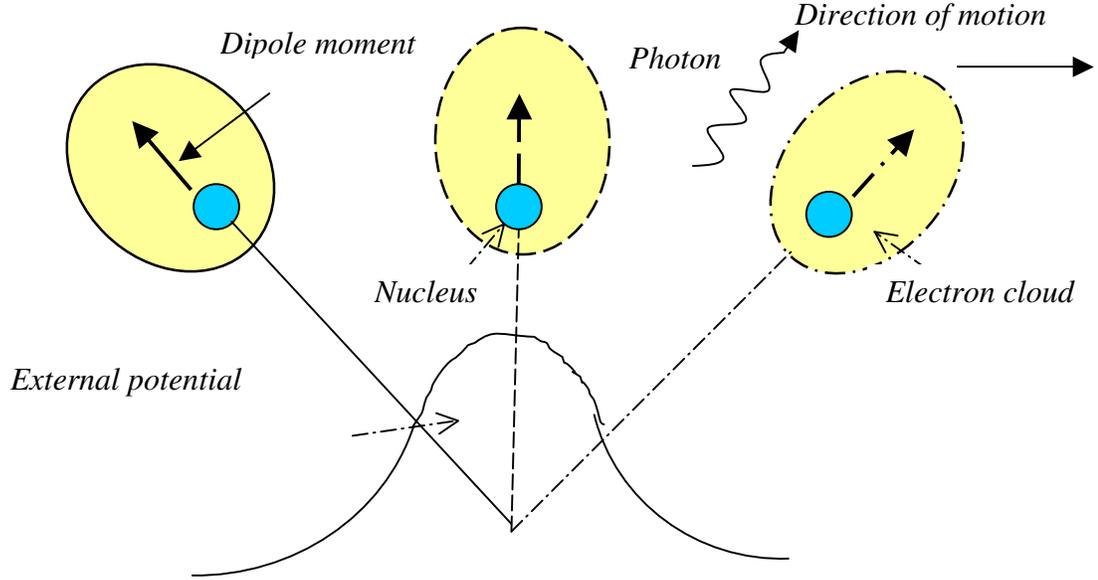

Fig. 2. Classical "atom" in an external field

As is shown in the Fig. 2, the induced dipole moment changes its orientation (rotates in space) and alters its magnitude, thus leading to photon emission. This mechanism could be of interest also in connection to radiation of not totally stripped ions in storage rings or ion accelerators. In these cases the deformation of electron distribution of the ions under the action of the accelerator's static field induces a dipole moment. The latter, changing in time its direction and magnitude, becomes a source of radiation.

In the cases presented by Fig. 1 and Fig. 2 the photon energy is taken from the projectile's kinetic energy, but the photon is emitted by the dipole moment induced in either the target (Fig. 1) or in the projectile, as in the case of Fig. 2 or when an atom instead of a simple charge is a projectile in "atom-atom" collision in Fig. 1

This article is devoted to consideration of PR (or AB) process, applied not to "classical" but to real atoms and ions and thus described by quantum instead of classical mechanics. However, the classical pictures present qualitatively accurate description of the AB process.

As to the atomic collisions, it was important to demonstrate not only the very fact of AB existence, but also the possibility that AB at least at some frequencies $\omega$ of the emitted photons could be considerable or even dominate over OB. The opposite view was supported by the widely known so-called *low energy theorem* formulated by Low (1958), according to which the BrS intensity I at emitted photon energy $\omega \to 0$ [1] can be presented as

$$I = \frac{A}{\omega} + B + C\omega, \qquad (1)$$

where A is totally determined by the OB contribution and B is expressible via the same characteristics of the projectile-target scattering process as A. As to AB, it contributes only to C and higher order terms in the expansion of I in powers of $\omega$.

---------

[1] Atomic system of units is used throughout this paper: the electron mass $m$, electron charge $e$ and Planck's constant $\hbar$ are equal to 1: $m = e = \hbar = 1$.

---------

However, for $\omega \geq I$, where $I$ denotes the characteristic excitation energy of the target, the intensity of AB, $I_{AB}$, is bigger than the intensity of OB, $I_{OB}$, as it was demonstrated in by Buimistrov and Trakhtenberg (1975) and by Amusia et al. (1976). In fact, detailed investigations demonstrated that $I_{AB}$ could considerably overcome $I_{OB}$.

Since the seventieth and then during eighties it appeared a lot of theoretical publications, the results of which were summarized in review articles and books [e.g. (Amusia, 1986, 1990), (Amusia et al., 1992), (Korol' and Solov'yov., 1997), (Korol et al., 2004)]. A number of specific features of AB, for example its complete independence of projectile mass were predicted. So, it seemed that the direct observation of AB is a simple matter. This proved to be not the case. During all this rather long period only one experiment directly and unambiguously confirming the existence of AB was performed by studying the radiation emission spectrum in collisions of medium-energy electrons with Xe gas target (Verkhovtseva et al., 1983). C. Quarles after a number of years of research, found noticeable contribution of PR to the total intensity of radiation by fast electrons going through gas targets (Portillo and Quarles, 2004).

From the very formulation of the AB mechanism its universal nature became evident. In other words, it became clear that AB leads to emission of radiation not only in electron – atom or ion – atom collisions, but in any case when at least one of the colliding particles, even being neutral, consists of electrical charges of different sign. As a result, time-dependent multiple moments are induced leading to emission of electromagnetic radiation. Of course, AB acts in molecular collisions and in collisions of macroscopic bodies that consist of atoms or ions and electrons. However, this mechanism acts also when microscopic projectiles like neutron or neutrino participate in a collision process. AB acts in their collisions with atom or nuclei. Moreover, since a neutron consists of at least three quarks, it can be also deformed in a collision with another neutron, thus generating emission of electromagnetic radiation. More general, mechanisms similar to AB lead to creation of other particles, like e.g. neutral $\pi$ - mesons, in nucleonic collisions.

**2. Radiation in electron – atom scattering**

In this Section we will investigate quantum-mechanically the most transparent and illustrative example of collisions, in which Atomic Bremsstrahlung is generated, namely the electron-atom collisions. The total BrS amplitude $F(\omega)$ can be presented as a sum of two contributions, having the first term $F^{OB}(\omega)$ that represents Ordinary Bremsstrahlung (OB) and the second term $F^{AB}(\omega)$ that stands for Atomic Bremsstrahlung (AB):

$$F(\omega) = F^{OB}(\omega) + F^{AB}(\omega). \qquad (1)$$

The bremsstrahlung amplitude of an incoming electron with the energy $E = p^2/2$ depends upon its momentum $\boldsymbol{p}$, the momentum $\boldsymbol{q}$ transferred to the target atom and the emitted photon polarization vector $\boldsymbol{e}$. At high but non-relativistic energies the electron-atom scattering amplitude looses its dependence upon $\boldsymbol{p}$, leading to a very simple and transparent expression for the BrS amplitude [2/] (see Amusia, 1990)

$$F_{\vec{p}\vec{q}}(\omega) = (\vec{e}\cdot\vec{q})\left[\frac{U(q)}{\omega} + \frac{4\pi\omega}{q^2}\alpha_d(\omega, q)\right]. \qquad (2)$$

Here $U(q)$ is the Fourier image of the atomic potential and $\alpha_d(\omega, q)$ is the generalized polarizability of the target atom.

---

[2/] Usually in this paper we will present simplified relations and equations in order to demonstrate qualitative features of a mechanism considered and a theoretical result obtained.

---

The BrS cross-section is expressed via the amplitude $F_{\vec{p}\vec{q}}(\omega)$:

$$\frac{d^2\sigma(\omega)}{d\omega dq} = \frac{\omega q}{16\pi^4 c^3 p^2} \int |F_{\vec{p}\vec{q}}(\omega)|^2 \, d\Omega_\gamma d\Omega_q, \tag{3}$$

where $d\Omega_{\gamma(q)}$ is the solid angle of photon emission (momentum $q$ transfer), $c$ is the speed of light.

At high $E$ the atomic potential becomes particularly simple both in coordinate and momentum spaces:

$$U(r) = -Z/r + \int d\vec{r}' \rho(\vec{r}')/|\vec{r}-\vec{r}'|,$$
$$U(q) = 4\pi[-Z + Q(q)]/q^2, \tag{4}$$

where Z is the nuclear charge, $\rho(r)$ and $Q(q)$ are the density of atomic electrons and its Fourier-image or so-called atomic form-factor. The combination $[Z - Q(q)]$ is the so-called *distributed* charge of the target atom. At $q \to 0$ one has $Q(0) = N$ and $\alpha_d(\omega,q) \to \alpha_d(\omega)$, where N is the total number of atomic electrons and $\alpha_d(\omega)$ is the dipole polarizability of the target atom. Atomic polarizability has poles at photon energies equal to the discrete excitation energies of the target atom. For medium and low energies $E$ the electron-atom potential is much more complex than (4): it is non-local and energy-dependent and cannot be presented in a close analytical form.

At $\omega \to 0$ the polarizability is positive, while at $\omega \to \infty$ it is given by the following relation $\alpha_d(\omega) \to -Q(q)/\omega^2$. As a result, at high $\omega$, which means photon energies much bigger than typical binding energies of atomic electrons, the BrS amplitude upon a neutral atom, $N = Z$, reduces to that upon a Coulomb field of a charge Z (see Akhiezer and Berestetsky, 1965):

$$F_{\vec{p}\vec{q}}(\omega) \to -4\pi(\vec{e}\vec{q})Z/\omega q^2. \tag{5}$$

This result has a simple qualitative explanation. Since for $\omega \gg I$ one can neglect the electron-nucleus binding, the atomic and the incoming electrons, as a system of free electrons cannot have a time-dependent dipole moment and therefore cannot radiate. What is left over is the radiation of the incoming electron in the Coulomb field of the nucleus (5). Amusia, Avdonina et al. (1985) called "atomic striptease" the elimination of electron's contribution to BrS at $\omega \gg I$. For $e + H$ this effect was demonstrated for the first time by Buimistrov and Trakhtenberg (1977).

In all textbooks BrS of an electron on an atom is described as OB of an electron in decelerating in the screened static field of atom (see e.g. Heitler, 1954). The inclusion of AB changes this result not only quantitatively, but qualitatively, since the ratio of OB to total BrS amplitudes is given for a neutral atom by the following expression:

$$F^{OB}_{\vec{p}\vec{q}}(\omega)/F_{\vec{p}\vec{q}}(\omega) = [Z - Q(q)]/Z \ll 1, \tag{6}$$

at least for small $q$. Thus, inclusion of AB corrects the generally

If the projectile is not an electron, but has instead a charge $e_i$ and mass $m_i$, the BrS amplitude is given by the following expression similar to (1)

$$F^{(i)}_{\vec{p}\vec{q}}(\omega) = e_i(\vec{e}\cdot\vec{q})\left[\frac{e_i U(q)}{m_i \omega} + \frac{4\pi\omega}{q^2}\alpha_d(\omega,q)\right]. \tag{7}$$

It is seen that if $m_i \gg 1$, the OB contribution can be neglected.

For neutral atoms to generate OB the projectile must come inside the target atom, while AB is generated by dipole, i.e. long-range interaction. Since AB and OB are created at long and short distances, respectively, their contributions to the total BrS add to each other almost without interference:

$$\frac{d\sigma}{d\omega} \approx \frac{16}{3}\frac{Z^2}{c^3 p^2 \omega}\ln 2pR_a + \frac{16}{3}\frac{\omega^3 |\alpha_d(\omega)|^2}{c^3 p^2}\ln\frac{p}{\omega R_a}, \qquad (8)$$

where $R_a$ is the target atom radius. Note, that at high $\omega$, where the relation $\alpha_d(\omega) \approx -Z/\omega^2$ is valid accurately enough, the formula (8) reduces to that for the radiation of an electron upon bare nuclei with the charge Z:

$$\frac{d\sigma}{d\omega} = \frac{16}{3}\frac{Z^2}{c^3 p^2 \omega}\ln 4E/\omega. \qquad (9)$$

The angular distribution of OB and AB in the considered collision is simple: it is similar to that of a rotating dipole. The angular dependence is given thus by the following expression (Amusia, Avdonina et al., 1985)

$$\frac{d^2\sigma}{d\omega d\Omega_\gamma} = \frac{1}{4\pi}\frac{d\sigma}{d\omega}[1 - \beta(\omega)P_2(\cos\theta_{\vec{k}\vec{p}})], \qquad (10)$$

where $P_2(\cos\theta_{\vec{k}\vec{p}})$ is the second order Legandre polynomial, $\theta_{\vec{k}\vec{p}}$ is the angle between vectors of the incoming electron and emitted photon momenta and $\beta(\omega)$ is the so-called radiation angular anisotropy parameter.

The fact that OB and AB are generated when the incoming electron is close to and far from the target atom respectively permits to obtain a rather simple expression for the amplitude of radiation absorption (RA) – a process that is time-reverse to BrS even at low incoming electron energies. The following relation gives this amplitude

$$F_{\vec{p}}^{RA}(\omega) \cong (\vec{e}\cdot\vec{p})\left[\frac{a}{\omega} + \alpha_d(\omega)\right]. \qquad (11)$$

Here $a$ is the so-called electron-atom scattering length, $p$ is the electron momenta after it absorbs a photon.. For a number of atoms the parameter $a$ is negative, e.g. in Ar, Kr and Xe. As a result, for these atoms the amplitude (11) reach zero at $a \approx -\omega_0 \alpha_d(0)$.

According to (8) and (11), the AB contribution is particularly big when the polarizability has its maxima – narrow in the region of discrete excitations of the target atom and broad and powerful that corresponds to Giant resonance's in the photoionization cross-section, which is typical for e.g. atoms with the $4d^{10}$ subshell, such as I, Xe, Cs, and Ba.

For excited atoms the polarizabilities are very big and presented by a maximum almost at the excitation frequency $\omega_{ex}$ with the width $\Gamma_{ex}$, thus rapidly decreasing at $\omega$ far from $\omega_{ex}$, when $|\omega - \omega_{ex}| \gg \Gamma_{ex}$. Therefore at $\omega \approx \omega_{ex}$ the BrS spectrum is strongly enhanced but at $\omega - \omega_{ex} \gg \Gamma_{ex}$ it rapidly approaches OB of an ionized atom, namely that, from which all electrons forming the excitation are eliminated.

The same is true also for alkali atoms, where the polarizability at low $\omega$, $\omega \sim I$, is determined by virtual excitations of the outer electron only. Therefore $\alpha_d(\omega)$ has big values concentrated in a narrow region of frequencies $\omega \sim I$. Radiation with $I_{inner} > \omega \gg I$ coincides with OB on alkali ion, without the outer electron. Here $I_{inner}$ is the inner subshell ionization potential of the alkali atom.

### 3. Relativistic collisions

In principle, the considered above AB mechanism is effective at any speed of the incoming electron. In the previous Section we discussed high energy, but non-relativistic electron - atom collisions. Relativism modifies the situation essentially, since the interaction between the incoming particle and the target atom as well as the very motion of the projectile is affected by relativism. Indeed, the relation between energy $E$ and momentum $p$ is not that simple any more as $E = p^2/2m_i$, becoming instead $E = \sqrt{c^2 p^2 + m_i^2 c^4}$ and the interaction between moving charges is not pure Coulomb. Its Fourier image becomes dependent not only upon the transferred momentum $q$, as $\sim 1/q^2$ but upon transferred frequency $\omega$ also, as $1/(q^2 - \omega^2/c^2)$. By choosing the so-called Coulomb calibration, it is possible to separate pure Coulomb inter-particle interaction that is determined by the exchange of so-called longitudinal photons and that determined by exchange of transverse photons. In the Coulomb calibration the interaction of a projectile with a charge motionless in the process of radiation is not affected by retardation. For $\omega \ll E \approx pc$ one has the following expression for the OB total spectrum of an electron in the Coulomb field of the charge Z (see e.g. Amusia, 1990):

$$\frac{d\sigma^{OB}}{d\omega} = \frac{16}{3} \frac{Z^2}{c^5 \omega} \ln[(2E/\omega)(E/c^2)]. \tag{12}$$

Comparing (12) with (9) we see that for OB relativism at $E \gg c^2$ adds a big factor $E/c^2 \gg 1$ under the logarithm. For a static field of a neutral atom one has instead of (12):

$$\frac{d\sigma^{OB}}{d\omega} = \frac{16}{3} \frac{Z^2}{c^5 \omega} \left( \ln cR_a - \frac{1}{2} \right). \tag{13}$$

It is essential that OB for a neutral atom in the relativistic case does not include the logarithmically increasing term, which characterize the high energy, but non-relativistic case (9)

The following relation gives the AB total spectrum (see Amusia et al., 1985a, Astapenko et al., 1985):

$$\frac{d\sigma^{AB}}{d\omega} = \frac{16}{3} \frac{Z^2 \omega^3}{c^5} |\alpha_d(\omega)|^2 \ln[(v/\omega R_a)(E/c^2)]. \tag{14}$$

Comparing (12) and the second term in (8), we see that the inclusion of exchange by transverse photons in the interaction of the projectile and target atom adds a big factor $E/c^2 \gg 1$ under the logarithm.

The BrS spectrum for relativistic and non-relativistic electrons differs considerably. At $v \to c$ and $\omega \gg I$ the mentioned in the previous section "atomic striptease" or de-screening is substituted by an opposite effect. Indeed, in the high $\omega$ limit, where $\alpha_d(\omega) \to -N/\omega^2$ (this is valid at $\omega \gg I$) one obtains by combining (13) and (14) the following expression

$$\frac{d\sigma}{d\omega} = \frac{16}{3} \frac{Z^2}{c^5 \omega} \ln E/\omega. \tag{15}$$

This formula differs only by a factor $Z^2$ from the spectrum of an ultra-relativistic electron upon an electron (Berestetskii et al., 1974). At $\omega \gg I$ the binding of atomic electrons to the nucleus can be neglected. The total BrS amplitude can be presented as a three-term sum:

$$F_{\vec{p}\vec{q}}(\omega) = Z F^{(1)}_{\vec{p}\vec{q}}(\omega) + Z F^{(2)}_{\vec{p}\vec{q}}(\omega) + F^{(3)}_{\vec{p}\vec{q}}(\omega), \tag{16}$$

where $F^{(1)}_{\vec{p}\vec{q}}(\omega)$, $F^{(2)}_{\vec{p}\vec{q}}(\omega)$, $F^{(3)}_{\vec{p}\vec{q}}(\omega)$ are amplitudes of projectile-atomic electron, atomic electron on projectile and projectile – nucleus photon emission, respectively. Since the BrS amplitude of the ultra-relativistic electron on a motionless free particle is independent upon its mass, being determined by its charge, the radiation amplitude upon atomic electrons and the nucleus, $Z F^{(1)}_{\vec{p}\vec{q}}(\omega) + F^{(3)}_{\vec{p}\vec{q}}(\omega) \approx 0$ and therefore $F_{\vec{p}\vec{q}}(\omega) \approx Z F^{(2)}_{\vec{p}\vec{q}}(\omega)$. It means that in the ultra-relativistic case the BrS spectrum is determined by the coherent radiation of N atomic electrons in the field of the incoming electron (Amusia et al., 1985a).

In the relativistic case the angular distribution of OB and AB are qualitatively different. Namely, the OB radiation is concentrated within a cone relative to the direction of the incoming electron momentum with an angle $\theta_{\vec{k}\vec{p}} < c^2/E < 1$. As to the AB radiation, its angular distribution is similar to that of non-relativistic case, since the rotation of the dipole induced in the target atom by the projectile proceeds with non-relativistic velocities. This is a consequence of relatively weak projectile – target Coulomb interaction that is unable to "bind" the induced dipole to the projectile strong enough in order to force it to rotate relativistic.

**4. Polarization Radiation in ion-atom collisions**

Let us start by considering the case when radiation is generated in collisions of a heavy charge, e.g. by a proton instead of electron, with an atom. The BrS amplitude in this case is determined by expression (7) with $m_i \equiv M \gg 1$. Differential in transferred by the heavy projectile energy $\omega$ and momentum $q$ BrS cross-section is given by (3) with $v^2$ substituting $p^2$. Since $m_i \gg 1$, AB dominates over OB up to very low $\omega$ and high $q$. The ratio $\eta$ of AB to OB amplitude can be easily presented in the following form:

$$\eta = \frac{4\pi\omega^2 \alpha_d(\omega)M}{e_i q^2 U(q)} \approx \frac{\omega^2}{Iq^2}\frac{N_p}{Ze_i} \sim \frac{\omega^2 M}{Iq^2}, \tag{17}$$

where an estimation for the polarizability $\alpha_d(\omega)$ is used $\alpha_d \approx N_p/I^2$, with $N_p$ being the number of electrons, contributing mainly to the atomic polarizability. It means that for a heavy charge AB is more important than OB up to very low frequencies, $\omega < I/\sqrt{M}$.

An expression that determines the BrS spectrum is similar to (8):

$$\frac{d\sigma}{d\omega} \approx \frac{16}{3}\frac{Z^2 e_i^4}{c^3 v^2 M^2 \omega}\ln 2vMR_a + \frac{16}{3}\frac{e_i^2 \omega^3 |\alpha_d(\omega)|^2}{c^3 v^2}\ln\frac{p}{\omega R_a}. \tag{18}$$

It is essential that while OB for heavy particles is suppressed by a factor $1/M^2 \ll 1$, the AB contribution is the same. As a result, e.g. for a proton, BrS is determined entirely by the AB mechanism. Since for $\omega \geq I$ AB in electron – atom collisions is bigger than OB, due to the polarization mechanism the BrS of a charge at such $\omega$ is almost independent on its mass.

Of course, we have assumed that in BrS process the target atom remains unaltered in the sense that its initial and final states remain the same. In fact, an electron or heavy charge scattering with an atom can be accompanied not only by radiation but also by target ionization and excitation. The ionized electrons can in their turn radiate, leading to so-called *inelastic* BrS, that differs from the considered above *elastic* BrS, in which the only result of projectile – target scattering is the emission of a photon. It is understandable however that in the inelastic BrS each atomic electron contributes individually, so the number of essentially participating electrons $N_p$ appears only as a factor in the inelastic cross-section. As to elastic AB, it includes the factor $N_p$ already in the amplitude, since $\alpha_d \sim N_p$. So, generally speaking, the ratio of elastic-to-inelastic AB spectrum is proportional to $N_p \gg 1$. However, at low frequencies the inelastic BrS spectrum $d\sigma^{in}/d\omega$ is even bigger than the elastic AB spectrum $d\sigma^{AB}/d\omega$, since at $\omega \to 0$ one has in accord with the Low theorem $d\sigma^{in}/d\omega \sim 1/\omega$.

Of special interest is the inelastic process, in which the incoming particle collides with an excited target and the emitted in the process photon carries away the excitation energy leaving the target in its ground state. The cross-section of such a process that can be considered as *Superelastic* (Landau and Lifshits, 1988) is non-zero, in fact even divergent as $1/p$ at zero projectile energy $E = p^2/M$.

Of special interest is the emission of radiation in the collision of two particles, both of which has internal structure, e.g. two atoms. Such a collision is a rather complicated process even without photon emission since it is accompanied by excitation or emission of one or several electrons. Here we will concentrate on photon emission process that proceeds without inelastic processes, such as excitation or ionization of the collision partners.

With some inessential for qualitative consideration restrictions the amplitude of BrS in atom – atom collisions can be presented (see in Amusia, 1990) in the following form

$$F_{\vec{p}\vec{q}}(\omega) = \frac{4\pi(\vec{e}\vec{q})}{q^2\omega}\left\{[Z_1 - Q^{(1)}(q)][Z_2 - Q^{(2)}(q)] \times \left(\frac{Z_2 - N_2}{M_2} - \frac{Z_1 - N_1}{M_1}\right) + \right. \\ \left. + \omega^2[\alpha_d^{(1)}(\omega,q)(Z_2 - Q^{(2)}(q)) - \alpha_d^{(2)}(\omega,q)(Z_1 - Q^{(1)}(q))]\right\}. \quad (19)$$

Here $Z_{1(2)}$, $M_{1(2)}$, $N_{1(2)}$, $Q^{(1(2))}(q)$, $\alpha_d^{(1(2))}(\omega,q)$ are the nuclear charge, mass, number of electrons, atomic form-factor and generalized polarizability of atoms 1 and 2, respectively. The expression (19) is a natural generalization of (2), with (4) taken into account. The upper line in (19) stands for OB while the second line represents the AB, namely the radiation of the first atom polarized by the distributed charge $(Z_2 - Q^{(2)}(q))$ of the second atom and the radiation of the latter polarized by the distributed charge $(Z_1 - Q^{(1)}(q))$ of the first atom. It is seen that for neutral atoms $Z_{1(2)} = N_{1(2)}$ the total radiation is AB. The formula (18) is correct at high collision velocities only. With decrease of velocity the corresponding qualitative picture of AB remains valid, however the radiation intensity cannot be expressed via polarizability of the collision partners. Instead, a sort of a common polarizability has to enter the BrS amplitude expression.

According to (19), for identical colliding particles, "1"="2", the BrS amplitude is zero. This fact has a simple qualitative explanation: identical charge distributions cannot form a time-dependent dipole moment and therefore cannot emit dipole radiation. However, in collisions of identical particles quadrupole radiation can be emitted. Its intensity for reasonably small $\omega$ is by a factor $(\omega R_a)^2 \ll 1$ smaller than the normal dipole radiation. It will be demonstrated in Section 10a that this fact plays a very important role in radiative cooling of volumes of gas.

The angular distribution of BrS generated in heavy particles collisions is, as it follows from its AB physical nature, the same as that of a rotating dipole, being proportional to $(1 + \cos^2\theta)$, where $\theta$ is the angle between the projectile and emitted photon momenta.

### 5. Relativistic ion-atom collisions

Relativistic ion-atom or atom-atom collisions are much more complex for theoretical treatment than the non-relativistic ones. We will see, however, that it is worthwhile to overcome the corresponding difficulties, since the emitted in collisions radiation has a number of peculiarities and interesting features. The complications in the relativistic collision of heavy particles have the same origin as in the case of electron – atom collision. These are consequences of relativistic expression for energy-momentum relation for each particle and of retardation in the interaction between moving charges. The interaction between a moving charge and a photon has to be also properly modified (see Berestetskii et al., 1974).

Qualitatively, the spectra of photons emitted in collisions of fast but non-relativistic and relativistic structured heavy particles, like ions or atoms, are essentially different because of Doppler effect and aberration that modify the frequency and the emission angle of a photon that originates from the object that moves relativistic.

The analytic expressions for the amplitude became considerably more complex than (19). Amusia et al. (1988) investigated the problem initially. It is more convenient to present it as a sum of two terms representing the photon emission by the projectile, $F^{(1)}_{\vec{p}\vec{q}_1}(\omega)$, and target, $F^{(2)}_{\vec{p}\vec{q}_1}(\omega)$, respectively:

$$F^{(1)}_{\vec{p}\vec{q}_1}(\omega) = \frac{4\pi(Z_1 - Q^{(1)}(q_1^{\perp}))}{(q_1^{\perp})^2}(\vec{e}\vec{q}_1)\frac{\omega}{c}\alpha_d^{(2)}(\omega, \vec{q}_1^{\perp}),$$

$$F^{(2)}_{\vec{p}\vec{q}_1}(\omega) = \frac{4\pi(Z_2 - Q^{(1)}(q_2^{\perp}))}{(q_2^{\perp})^2}\left[\frac{\tilde{\omega}}{c}(\vec{e}\vec{q}_2) + \gamma\beta(\vec{e}n_1)(\vec{k}\vec{q}_2^{\perp})\right]\alpha_d^{(1)}(\tilde{\omega}, \vec{q}_1^{\perp}),$$
(20)

where $\tilde{\omega} = \omega\gamma(1-\beta\cos\theta)$, $\beta = v/c$, $\gamma = (1-\beta^2)^{-1/2}$, $\theta$ is the vector between momenta **k** and **p**$_1$, $\vec{q}_1^{\perp} \approx -\vec{q}_2^{\perp}$. In deriving (20) it was taken into account (see Amusia, 1990) that in the total transferred momentum $\vec{q} = \vec{q}^{\parallel} + \vec{q}^{\perp}$ only the longitudinal component $\vec{q}^{\parallel}$ (parallel to $\vec{p}_1$) of the projectile momentum is modified when going from the projectile to target coordinate frame. It is essential that since $\omega = q^{\parallel}c$ and we consider not too high radiation frequencies $\omega << cR_a^{-1}$ while for colliding of neutral atoms $q \sim R_a^{-1}$, one has $q_1^{\parallel} << q_1^{\perp}$.

Using the amplitude (20), one can obtain differential in $\omega$ and $\theta$ expressions for the AB cross-sections of the target (*t*), projectile (*p*), and the corresponding interference term (*i*):

$$\frac{d^2\sigma^{(t)}(\omega)}{d\omega d\Omega_\gamma} = \frac{\omega^3}{\pi v^2 c^3}(1+\cos^2\theta)\int_0^\infty \frac{dq^{\perp}}{q^{\perp}}\left|(Z_1 - Q^{(1)}(q^{\perp}))\alpha_d^{(2)}(\omega, q^{\perp})\right|^2,$$

$$\frac{d^2\sigma^{(p)}(\omega)}{d\omega d\Omega_\gamma} = \frac{\omega\tilde{\omega}^2}{\pi v^2 c^3}\left[1+\left(\frac{\cos\theta-\beta}{1-\beta\cos\theta}\right)^2\right]\int_0^\infty \frac{dq^{\perp}}{q^{\perp}}\left|(Z_2 - Q^{(2)}(q^{\perp}))\alpha_d^{(1)}(\tilde{\omega}, q^{\perp})\right|^2, \quad (21)$$

$$\frac{d^2\sigma^{(i)}(\omega)}{d\omega d\Omega_\gamma} = -\frac{2\omega^2\tilde{\omega}}{\pi v^2 c^3}\left[1+\cos\theta\left(\frac{\cos\theta-\beta}{1-\beta\cos\theta}\right)\right]\int_0^\infty \frac{dq^{\perp}}{q^{\perp}}\text{Re}[(Z_1 - Q^{(1)}(q^{\perp}))(Z_2 - Q^{(2)}(q^{\perp}))\times$$
$$\alpha_d^{(1)}(\tilde{\omega}, q^{\perp})(\alpha_d^{(2)}(\omega, q^{\perp}))^*]$$

Comparing the radiation of the target and projectile, we see that the projectile (*p*) radiation is affected in the target frame by angular aberration that leads to substitution $(1+\cos^2\theta)$ by $\{1+[(\cos\theta-\beta)/(1-\beta\cos\theta)]^2\}$ and by the Doppler effect that alters the frequency: $\omega \to \tilde{\omega}$. Because of this, a narrow resonance line in the polarizability $\alpha_d^{(1)}(\omega, q^{\perp})$ at $\omega = \omega_0$ in the projectile frame transforms into a broad band $\omega_0\gamma^{-1} < \omega < \omega_0\gamma$ in the laboratory frame. Thus, it is seen that even relatively small frequencies in the projectile frame can be increased by a factor up to $\gamma$.

The relative contribution of projectile and target radiation depends upon the collision velocity. Indeed, as is seen from (21), if the following inequality is fulfilled

$$\gamma^2(1-\beta\cos\theta)^2 |\alpha_d^{(1)}(\omega\gamma(1-\beta\cos\theta))|^2 >> |\alpha_d^{(2)}(\omega)|^2, \quad (22)$$

then the projectile radiation dominates. If an opposite inequality holds, the target becomes the main source of radiation.

As is seen from (21), the angular distribution of the target radiation is like that of a rotating dipole. The angular distribution for the projectile radiation in relativistic atom-atom collisions depends upon $\omega$. Being in general rather complex, it is simplified in two limiting cases of very low and high frequencies $\omega$, $\omega << I/\gamma$ and $\omega >> I\gamma$, respectively. In the first case the angular dependence is proportional to $[(1-\beta\cos\theta)^2 + (\cos\theta - \beta)^2]$ instead of being concentrated within a narrow cone with an apex angle $\theta \leq \gamma^{-1}$, as it is for a charged projectile. In the second case the projectile generates radiation just as a charge. This is understandable, since for $\omega \gg I\gamma$ all projectile electrons can be treated as free charges radiating coherently in the field of the target.

According to (19), two identical particles do not radiate. The situation in relativistic collisions is essentially different: the radiation intensity is non-zero, as it follows from (21). Qualitatively, the complete compensation of dipole moments induced in the projectile and target disappears because in e.g. the target frame the projectile's induced moment is shifted in size and direction relative to that of the target due to Doppler effect and angular aberration.

Corresponding expressions can be obtained from (21). We will present them for the case $\sqrt{I} \ll v \ll c$, where the obtained radiation appears as a relativistic correction to the zero non-relativistic value that follows from (19) for $(1) = (2)$. For $\omega \ll I/\gamma$ and $\omega \gg I\gamma$, one has, respectively

$$\frac{d\sigma^{(id)}}{d\omega} = \frac{32}{3}\frac{\omega^3}{c^5}\int_0^\infty \frac{dq^\perp}{q^\perp}[Z-Q(q^\perp)]^2 \alpha_d^2(0,q^\perp),$$

$$\frac{d\sigma^{(id)}}{d\omega} = \frac{16}{5}\frac{1}{c^5\omega}\int_0^\infty \frac{dq^\perp}{q^\perp}[Z-Q(q^\perp)]^2 Q^2(q^\perp). \tag{23}$$

Comparing (23) with (8) and (9), one concludes that radiation of identical particles is by a factor of $(v/c)^2$ smaller than that of different particles. Note that in all the above expressions in this Section it is implied that $q^\perp = \sqrt{q^2 - q^{\|2}}$, where $q^\|$ is fixed by the relation $q^\| = \omega/c$.

### 6. Collisions with quasi-atomic objects - clusters and fullerenes

It was demonstrated in the previous sections that AB is totally determined by the generalized polarizabilities $\alpha_d(\omega,q)$ of the colliding objects. The bigger are the polarizabilities, the more important is the AB contribution. Relatively recently multiatomic objects were created that have extremely big polarizabilities in quite broad frequency regions, namely metallic clusters and fullerenes, the most well known of which is the molecule $C_{60}$. The polarizability of metallic clusters can reach tens of thousands of atomic units, as compared to the atomic value of about unit for atoms. The generalized polarizability $\alpha_d(\omega,q)$ for $C_{60}$ is by two orders of magnitude bigger than that for atoms.

The photoionization cross-sections of fullerenes and metallic clusters is characterized by Giant resonances, similar but much more powerful than that for $4d^{10}$ electrons of Xe. The total oscillator strength of this resonance in Xe is about ten, while in $C_{60}$ it is by more than an order of magnitude bigger, of about two hundred. In metallic clusters the total oscillator strength of Giant resonance's is tremendous, reaching hundred thousands.

To estimate the AB spectrum one can use expression (8), substituting there atomic polarizability $\alpha_A(\omega)$ by that of fullerenes or clusters $\alpha_F(\omega)$, as it was done for $C_{60}$ by Amusia and Korol'(1994). Roughly estimating $\alpha_F(\omega)$ as $\alpha_F(\omega) \approx N\alpha(\omega)$, one obtains

$$\left(\omega\frac{d\sigma}{d\omega}\right)_{AB} = \frac{16N^2}{3v^2c^3}\omega^4|\alpha_A(\omega)|^2 \ln(v/\omega R), \tag{24}$$

where $R$ is the radius of the multiatomic object.

The OB contribution has to be modified as compared to (8). Indeed, OB originates from contributions of isolated atoms

$$\left(\omega\frac{d\sigma}{d\omega}\right)_{OB} = \frac{16}{3v^2c^3}\int_{\omega/v}^{2v}\frac{dq}{q}n^2(q)[Z-Q_A(q)]^2, \qquad (25)$$

where $n(q)$ is the Fourier transform of the density of individual atoms distribution in the fullerenes or cluster. The function $n(q)$ is concentrated at small $q$-values, $q < R^{-1}$, while the main contribution from $[Z-Q_A(q)]$ is in the $q > R_A^{-1}$ region (where $Z$ is the constituent atom's nuclear charge), $R_A$ is the radius and $Q_A(q)$ is the atomic form-factor, defined by (4)). Since $R >> R_A$ there is no overlap of $n(q)$ with that part of $q$-space where $[Z-Q_A(q)]$ is concentrated. Estimating $n(q)$ for $q \geq R_A^{-1}$ as $n(q) \approx N^2/2(qR)^2$ (where $N$ is the total number of atoms in the multiatomic system) one obtains from (24) for fullerenes, where atoms are located at its surface, so that $(R_A/R)^2 \sim N^{-1}$, the following relation:

$$\left(\omega\frac{d\sigma}{d\omega}\right)_{OB} \approx \frac{4N^2}{3v^2c^3}Z^2(R_A/R)^2 \approx \frac{16N}{3v^2c^3}Z^2. \qquad (26)$$

Therefore, due to absence of a concentrated nucleus in clusters and fullerenes, OB at $\omega \geq I$ is suppressed relative to OB at these $\omega$ by a factor $N^{-1} << 1$, i.e. much stronger than in atoms. As a result AB dominates completely in the BrS spectrum.

Giant resonances in photoabsorption cross-sections $\sigma_\gamma(\omega)$ of atoms, fullerenes and metallic clusters are represented by almost symmetric maxima. Therefore for real and imaginary parts of the polarizabilities one has $\text{Re}\,\alpha(\omega) << \text{Im}\,\alpha(\omega) = c\sigma_\gamma(\omega)/4\pi\omega$, which results instead of (26) in the following expression:

$$\left(\omega\frac{d\sigma}{d\omega}\right)_{AB} = \frac{\omega^2\sigma_\gamma^2(\omega)}{3\pi^2v^2c^3}\ln(v/\omega R), \qquad (27)$$

Estimating $\sigma_\gamma(\omega)$ at the resonance frequency $\omega_{max}$ as $\sigma_\gamma(\omega_{max}) \sim N_e/\Gamma$, where $N_e$ is the number of collectivized electrons in the radiating object and $\Gamma$ is the Giant resonance's width, one obtains for the ratio $\eta = \frac{(\omega d\sigma/d\omega)_F}{(\omega d\sigma/d\omega)_A}$ the following expression:

$$\eta = \frac{(\omega d\sigma/d\omega)_F}{(\omega d\sigma/d\omega)_A} = \left(\frac{N_{e,F}}{N_{e,A}}\right)^2\left(\frac{\omega_{max,F}}{\omega_{max,A}}\right)^2\left(\frac{\Gamma_A}{\Gamma_F}\right)^2. \qquad (28)$$

Substituting into (28) the numbers for $C_{60}$ and $4d^{10}$ of Xe, respectively, one has $\eta \approx 200$, which means a tremendous enhancement, having in mind that AB in the $4d^{10}$ Giant resonance region in Xe is quite prominent and was observed clearly by Verkhovtseva et al. in 1983.

Presented above consideration of BrS in electron-cluster or electron-fullerene collisions is satisfactory for estimations only. Having in mind that polarizability of these targets is very big, the action upon the incoming and outgoing electrons of corresponding polarization potential may be essential. Big polarizability can affect the transverse photon propagation leading to essential contribution of these photons to projectile – target electron interaction even at $\omega << c^2$. It is essential to note that in electron – fullerenes collisions the non-dipole radiation becomes considerably more important than in the atomic case, since the radiating objects are much bigger than atoms.

Of special interest is the BrS of clusters, in which electrons of individual atoms are not collectivized, as in metallic clusters or fullerenes. We believe, that in this case AB is also enhanced, as compared to isolated atoms. Indeed, AB is generated at big distances, of the order of $r_g \sim v/\omega$. As soon as $r_g$ becomes bigger than the inter-atomic distances $r_{int}$, $r_{int} > r_g$, instead of a single several target atoms become polarized simultaneously, thus leading to effective polarizability $\alpha_{eff}(\omega) \approx (r_g/r_{ia})^3 \alpha_A(\omega)$. The AB intensity is proportional to $|\alpha_{eff}(\omega)|^2$ thus acquiring an additional factor leading $(r_g/r_{ia})^3 \sim n(v/\omega)^3$, where $n$ is the density of atoms in the cluster.

There exists a whole variety of fullerenes, of which $C_{60}$ was selected as the best known. Other fullerenes could be also a source of radiation. If fullerenes or clusters have a non-spherical form, they can have not one, but two Giant resonances and AB would be represented by two maxima, correspondingly.

### 7. Inelastic Polarization Radiation

An important specific feature of collisions with atoms and atom-like objects is their ability to be ionized or disintegrated, which leads to emission of secondary electrons. Moving in the field of residual ions, these electrons emit OB. So, entirely, the total radiation created in the considered collisions is a sum of normal OB, AB and radiation of secondary electrons that consist of what could be called secondary or inelastic OB and AB or OB$_{in}$ and AB$_{in}$.

To calculate such a spectrum is a challenge indeed. Amusia et al., 1987 investigated this problem for the first time. For high $\omega$ and $q \leq 1/R_{at}$ the inelastic BrS dominates parametrically over inelastic one. Indeed, the summation over contribution of atomic electrons in elastic BrS is performed in the amplitude, which is manifested in the fact that $\alpha_d(\omega,q) \sim N_e$, with $N_e$ being the number of actively participating in AB electrons, enters the amplitude (2). As a result the elastic spectrum is proportional to $(d\sigma/d\omega)_e \sim N^2$. In inelastic scattering each participating electron generates BrS independently, so that $(d\sigma/d\omega)_i \sim N$ and if $N \gg 1$ $(d\sigma/d\omega)_e \gg (d\sigma/d\omega)_i$. Amusia et al., 1987 presented the total inelastic BrS cross-section as

$$\left(\frac{d\sigma}{d\omega}\right)_{in} = \frac{16N}{3c^3 p^2 \omega} \ln 2pR_a. \tag{29}$$

This contribution is by a factor of $N$ smaller than the first term in (8) that represent the elastic OB contribution.

The situation for low $\omega$ becomes considerably different. Elastic and inelastic OB are both $\sim 1/\omega$, while AB contribution is much smaller, being of the order of $\sim \omega$.

Inelastic BrS could be of great importance in cluster-cluster or fullerenes-fullerene collisions, since these are objects that can be easily ionized or even entirely disintegrated. However since the number of collectivized electrons $N$ is very big for these objects, the elastic AB dominates strongly over all the inelastic BrS.

Entirely, it was demonstrated that elastic AB dominates over OB, OB$_{in}$ and AB$_{in}$ in the following $\omega$ region, $I \leq \omega \leq vR_a^{-1}$. This fact is essential from the point of view of the AB experimental detection: in this $\omega$ region all the BrS spectrum comes from the elastic AB.

Of interest is a specific example of inelastic BrS, in which the incoming electron collides with an exited atom. Corresponding BrS that is accompanied by the target's de-excitation can be called *super elastic* BrS. If the projectile is relativistic, one has to take into account exchange of real instead of only virtual photons between the excited target and incoming particle. As a result, this process becomes resonant and its amplitude can be presented as a product of photon emission by the excited target and Compton scattering on the projectile (Batkin and Almaliev, 1985).

**8. Nucleon-atom collisions**

To describe the main contribution to the radiation spectrum generated in proton-atom collisions one can use (18) putting there $Z = 1$ and $M = M_p$, where $M_p$ is the proton mass. Radiation in neutron - atom collisions requires, however, special consideration, which was carried out for the first time by Amusia et al., 1987 (see also Amusia et al., 1992). Let us at first neglect the only direct neutron-electron interaction, namely the magnetic one. Then the incoming neutron interacts only with the nucleus. As a result of this interaction the nucleus is experiencing recoil and became shifted off its previous position right in the atomic center. This shift being time-dependent induces an alternative dipole moment in the surrounding electron shells and thus generates AB. The corresponding differential in $\omega$ cross section is given by

$$\frac{d\sigma}{d\omega} = \frac{16\omega^3}{3c^3}\left|f_E^n \alpha_d(\omega)\right|^2 \frac{E}{M_n}\sqrt{1-\frac{\omega}{E}}\left(2-\frac{\omega}{E}\right), \tag{30}$$

which is valid if the inequalities $p = \sqrt{2M_n E} \ll M_n/2$ and $\omega \leq E = M_N M_n v^2 / 2(M_n + M_N)$ are satisfied. In (30) $f_E^n$ denotes the neutron-nucleus scattering amplitude, $M_n$ and $M_N$ stand for the neutron and nucleus masses, respectively. As a function of $E$, $f_E^n$ for a given nucleus can have narrow powerful so-called Feshbach resonances. But by the order of magnitude in the non-resonant area $f_E^n \sim R_N$, where $R_N$ is the nucleus radius.

It is reflected in (30) that the neutron-atom AB amplitude has two scales, namely the nuclear one that represents the neutron-nucleus scattering at about $10^{-13} \div 10^{-12}\,cm$ and the atomic at $10^{-8}\,cm$. Obviously, the neutron radiation cross section is much smaller than that of a proton, where AB is of the same order as in the electron-atom scattering. This smallness is seen from (30), where $1/M_n \approx 1/2000$ and $f_E^n \sim R_N \sim 10^{-4} \div 10^{-5}\,at.un.$

Contrary to the short-range neutron - nucleus interaction, neutron's magnetic moment creates a long-range field that is proportional to $1/r^2$ (where $r$ is the neutron – electron distance) that polarizes the target atom, thus generating AB. To take into account the interaction between neutron's magnetic moment and atomic electrons, one has to add the following potential:

$$V = -\frac{\vec{\mu}}{M_n c}\left[\vec{E} \times \hat{\vec{p}}\right] - \mu\vec{H}. \tag{31}$$

This potential describes the interaction of neutron's magnetic moment with the electric *E* and magnetic *H* atomic fields, respectively. The first term in (31) describes the Schwinger's neutron scattering in the electric field of an atom, while the second represents the neutron scattering in the magnetic field, created mainly by the orbital moments of atomic electrons. OB of neutron's magnetic moment is negligible as compared to respective AB, since $M_n \gg 1$.

Particularly simple is the neutron's magnetic moment contribution for very slow neutrons, $p \ll 1$ (or $E \ll M_n^{-1}$). In this domain the AB cross section is given by the following expression (see Amusia et al., 1992):

$$\frac{d\sigma^{(\mu)}}{d\omega} = \frac{16\omega^3}{3c^3 v^2}\left|\alpha_d(\omega)\right|^2 \mu^2\left[\left(\frac{\omega}{c}\right)^2 \ln\frac{1+\sqrt{1-\omega/E}}{1-\sqrt{1-\omega/E}} + 4\left(\frac{E}{\omega}\right)^2 \sqrt{1-\frac{\omega}{E}}\left(1-\frac{\omega}{2E}\right)\right]. \tag{32}$$

It is of interest to compare the relative contributions of magnetic moment and recoil contributions to the AB in neutron-atom collisions, that are given by (31) and (30), respectively. From these expressions one has for the ratio $\xi_\mu \equiv d\sigma^{(\mu)}/d\sigma$:

$$\frac{d\sigma^{(\mu)}}{d\sigma} \sim \frac{\mu^2(E/c)^2}{(f_E^n)^2(E/M_N)^2} \approx \left(\frac{M_N}{M_n}\frac{1}{f_0^n c^2}\right)^2. \tag{33}$$

In the estimation (33) the known value of the neutron's magnetic moment $\mu \cong 1.91(2M_n c)^{-1}$ is used. Assuming that $f_0^n \sim R_N$ and $R_N \sim A^{1/3}$, one has $\xi_\mu` \sim A^{4/3}$, where $A$ is the total number of nucleons in the nucleus. Naturally for hydrogen one has $N = 1$. Using known numerical values, it is $\xi_\mu^{(1)} \approx 4 \cdot 10^{-2}$ for hydrogen and $\xi_\mu^{(200)} \approx 50$ for a nucleus with $A = 200$. This dramatic growth of $\xi_\mu` \sim A^{4/3}$ is a result of increase of magnetic neutron – electron interaction and decrease of recoil with $A$.

### 9. Nucleon-nucleus collisions

When a nucleon collides with a nucleus, the latter becomes deformed and emits polarization radiation (or AB) with $\omega$ of the order of nucleon binding energies in nucleus. The radiation is strongly enhanced in the region of nuclear Giant resonance energies (see e.g. Amusia et al., 1992) similar to what happens in atomic collisions. The essential difference is however in the fact that the quadrupole radiation in nuclei are much more pronounced than in atoms.

The amplitude $F_{\vec{p}\vec{q}}^n(\omega)$ of BrS generated in proton – nucleus collision is given by an expression similar to (2):

$$F_{\vec{p}\vec{q}}^p(\omega) = (\vec{e}\cdot\vec{q})\left[\frac{U_{pN}(q)}{\omega} + u_{pp}(q)\alpha_N(\omega,q)e_p + u_{pn}(q)\alpha_N(\omega,q)e_n\right]. \tag{34}$$

Here $U_{pN}(q)$, $u_{pp}(q)$ and $u_{pn}(q)$ are the Fourier images of the proton-nucleus, proton-proton proton – neutron potentials, respectively; $\alpha_N(\omega,q)$ is the nucleus generalized polarizability. To enhance similarity with the case of radiation in electron – atom collisions, equation (34) is written in the "nuclear" system of units, $e = \hbar = M = 1$ where $M$ is the nucleon's mass. The effective proton and neutron electrical charges, $e_p$ and $e_n$, depend upon the multipolarity of the emitted $\gamma$-quantum: for dipole quanta $e_p \approx N/A$, while $e_n \approx -Z/A$, where $Z$ and $N$ are the numbers of protons and neutrons in a nucleus, $Z + N = A$. For higher multipoles one has $e_p = 1$, $e_n = 0$. The existence of the effective charge in nucleus is a consequence of the fact that e.g. dipole excitation is only the movement of protons relative to neutrons while the movement of the nucleus center of gravity is recoil, not an excitation (see e.g. Bohr and Mottelson, 1973).

For an incoming neutron, the first term in (34) disappears, and radiation is entirely of polarization nature if $A \gg 1$, so that one can neglect the recoil contribution.

An incoming nucleon with an energy from several hundreds MeV to 1 GeV, interacts almost equally with a nuclear proton and neutron. Therefore the dipole excitation is suppressed and most essential become the quadrupole excitations. Dipole photons are coming from the first term in (34), while quadrupole from the second. Because of important contribution of the quadrupole radiation the angular distribution is different from that in the electron-atom case (10), acquiring an additional important contribution proportional to $P_4(\cos\theta_{\vec{k}\vec{p}})$. Concrete calculations performed in (Amusia et al., 1984) demonstrated that OB dominates in the nuclear Giant resonance's (see Bohr and Mottelson, 1973) $\gamma$- ray energy region.

"Striptease" in nuclear collisions is important. It manifests itself clearly in the radiation that is generated in neutron–deuteron collisions. This case illustrates also the prominent role played by recoil. Using the zero range approximation for the nucleon – nucleon interaction, which is definitely satisfactorily for estimations, one has instead of (7) the following expression

$$F_{\vec{p}\vec{q}}^{(nD)}(\omega) = \frac{4\pi(\vec{e}\cdot\vec{q})}{\omega}\left[(f_{np}+f_{nn})Q^{(D)}(q/2) + \omega^2(f_{nn}-f_{np})\alpha_d^{(D)}(\omega,q)\right] \quad (35)$$

Here $f_{nn}$ and $f_{np}$ are neutron-neutron and neutron-proton scattering amplitudes, respectively; $Q^{(D)}(q/2)$ and $\alpha_d^{(D)}(\omega,q)$ are the deuteron's form-factor and generalized polarizabilities, respectively. At $\omega \gg I$ $\alpha_d^{(D)}(\omega,q) \to -Q^{(D)}(q/2)/\omega^2$ and $F_{\vec{p}\vec{q}}^{(nD)}(\omega)$ reduces to

$$F_{\vec{p}\vec{q}}^{(nD)}(\omega) = 8\pi(\vec{e}\cdot\vec{q})f_{np}Q^{(D)}(q/2)/\omega \ . \quad (36)$$

The elimination of $f_{nn}$ from $F_{\vec{p}\vec{q}}^{(nD)}(\omega)$ illustrates the "striptease" in neutron-deuteron collisions similar to that in $e+H$ collisions. Calculations have demonstrated that for the neutron-deuteron case PB contributes up to 30-50% of the total BrS cross section.

Similarly to BrS in nucleon-atom and nucleon-nucleus collisions, the generation of radiation in meson-atom and meson-nucleus collisions can be considered.

**10. Polarization radiation as a general phenomenon**

In the previous Sections we have presented what can be considered to large extent as the *retrospective* and *current* ideological *status* of the AB. Here in this Section we
discuss a number of processes, including those that take place in Nature, where AB obviously persists but its real role is far from being sufficiently clear. The investigation of AB under discussed below conditions presents a part of what I see as *perspectives* in this field.

a. *Radiation in media and enhancement mechanisms*

Above in this paper we have discussed elementary processes, where AB is created in collisions of a single projectile with a single target. The proximity of several atoms in the target or electrons/atoms in the projectile can modify AB considerably. However, if the atomic density of the target is high enough, an incoming electron interacts with a number of atoms simultaneously, leading to serious modification of the media effective polarizability, which is illustrated by Fig. 3:

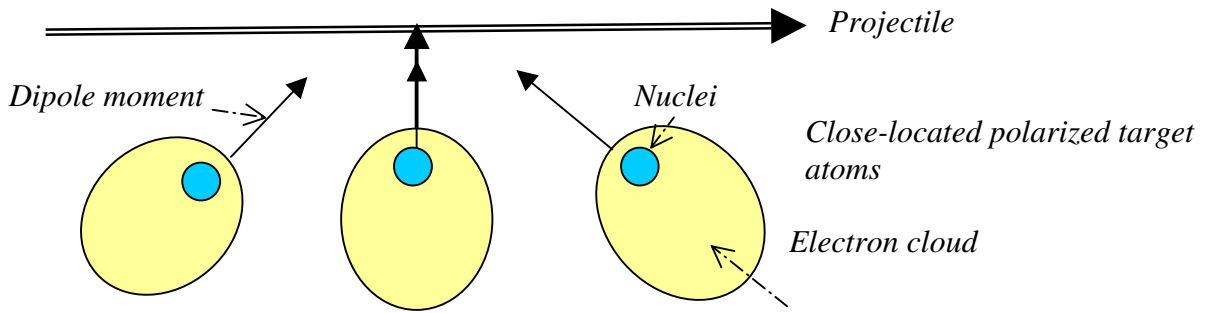

Fig. 3. Formation of a common induced dipole moment by a single projectile

If an incoming particle moves in amorphous media, the dipole momenta induced in different targets atom can effectively compensate each other, since on the average to each atom located at a given distance on one side of the projectile trajectory, thus totally compensating the dipole momenta induced in each atom, as is illustrated in Fig. 4. It is seen that dipole moments $\vec{D}_i(\omega)$ of four deformed atoms located symmetrically relative to the projectile can completely compensate each other:

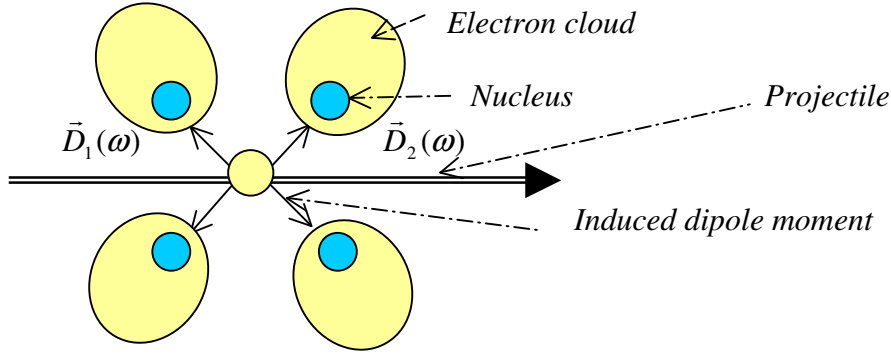

Fig. 4. Suppression of induced dipole moments in a symmetric environment.

Random distribution of target atoms guarantees good averaging of the induced dipole moment, leading for long enough wavelengths of the radiation to almost zero AB values. However, if the target has a crystalline structure, there are directions and positions of trajectories inside the crystal, following which the AB is enhanced, while when the projectile moves along other directions AB is suppressed. AB for crystal or amorphous media is a very interesting subject, which is not yet investigated sufficiently well.

Essentially different becomes the situation, when a projectile moves parallel to surface of a crystalline body. In this case AB is enhanced as compared to that for an isolated atom, if interatomic distance $r_g$ satisfies the relation $r_g \sim v/\omega$.

If an atom moves parallel to a surface, the AB of the projectile as well as the target can be generated due to action of the projectile upon deformable surface potential and vice versa. Of interest is the movement of a relativistic particle near a ferromagnetic or anti-ferromagnetic surface. In the projectile coordinate frame the periodic magnetic field of the surface is transformed into periodic electrical potential that act opposite on the projectile nucleus and its electrons, thus inducing AB in the projectile.

If a surface has a long-range structure with periodicity $a$ (e.g. due to its sound long wave oscillations), a relativistic projectile in it's own coordinate frame due to Lorenz contraction feels a field with the period $a/\gamma$ that can be of the projectile size, $R_p \sim a/\gamma \ll a$, and generates AB.

b. *Radiation in the processes of channeling*

Channeling is a well-known process of radiation by a particle, in the first place a positron, that moves along one of the crystal axis. A good model of channeling is a potential tube, in which an incoming positron is moving under a small angle to the tube's channel. Bouncing inside the tube, being repulsed from one side of the potential wall to another, the positron radiates. Since its speed is very close to that of light, the radiation is concentrated inside a very narrow cone near the axis, with an apex angle $\theta < c^2/E \ll 1$. The intensity of this directed radiation that can be considered in principle as OB is very high. However along with this radiation, AB can be essential in this process.

Indeed, the wall of the channel is not structureless and rigid, but instead consists of atoms (or ions) of the crystal. When the positron approaches the atom of the wall, it became polarized and emits AB. As it was discussed in the previous subsection, it is possible that not a singe atom from the channels wall, but a group of them is polarized simultaneously. This enhances the AB of the wall.

c. *Radiation by intensive currents*

Till now we have investigated radiation caused by a single projectile: an electron, positron, atom or ion. However, in principle of interest is the situation, when the beam of projectiles is so dense that simultaneously not one but many electrons interact with a single target object: atom, fullerenes, cluster. Schematic picture of this possibility is depicted in Fig. 5:

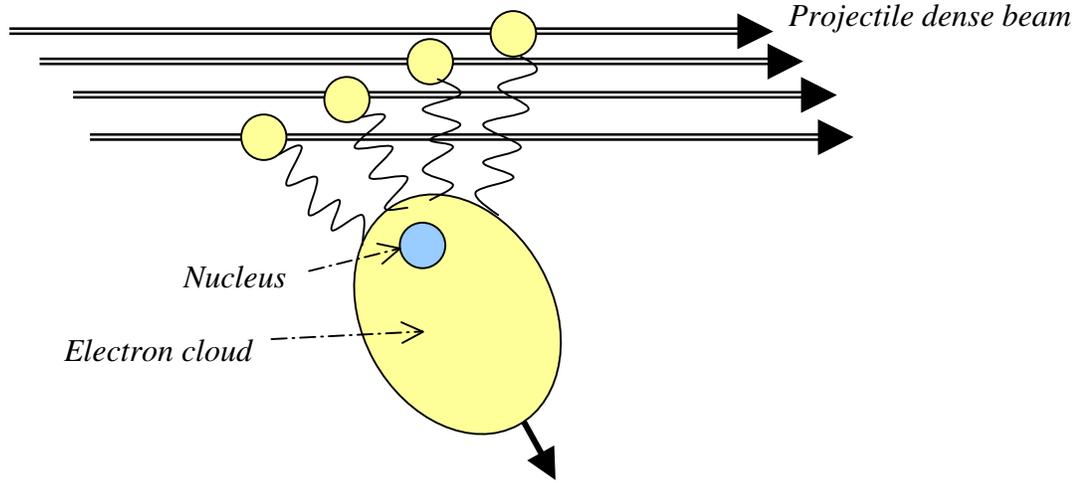

Fig. 5. Several projectiles interact simultaneously with one atom

The radiation in this case has to be proportional to the projectile density. It is evident that a sufficiently dense beam of same charge particles is hard to create due to their mutual Coulomb repulsion. This repulsion can be compensated by magnetic force, generated by the beam's own current. Roughly speaking, the total number of electrons, participating simultaneously in AB generation $N_{AB}$ is given by the characteristic volume of AB creation, $\sim (r_g)^3 \sim (v/\omega)^3$ multiplied by the beam's density $n_b$, so that $N_{AB} \sim n_b (v/\omega)^3$. The total AB intensity has to be proportional to $(N_{AB})^2 \sim [n_b (v/\omega)^3]^2$, since $N_{AB}$ is the charge polarizing the target atoms, clusters or fullerenes.

d. *Mechanical origin of radiation*: *sound and shock waves, cracking*

AB is generated always when positive or negative charges are moved in the collision process relative to each other, thus dipolar polarizing the target. Until now we have discussed only polarization caused by incoming electrons, positrons or atoms. However, polarization can be caused by mechanical action as well. Indeed, if accelerated, an atom can become polarized since mechanical action can be different upon atomic nucleus and electrons. One should keep in mind that from the microscopic point of view mechanical motion (and action) of a body is a cooperative motion (action) of atoms that form the body. Therefore, it is natural that if this motion proceeds with acceleration, or, better to say, with time-dependent acceleration, AB is generated. Propagation of ordinary sound serves as a first example of such a process. Being an oscillation of density of the media, a sound wave has to be accompanied by extra atomic collisions as compared to their number and intensity in the equilibrium state, thus generating AB. As it will be discussed in Section 12a, the AB spectrum generated in gas depends importantly on whether this gas consist of one sort of atoms or two and more: the radiation in the first case is at least by four orders of magnitude less intensive than in the second case. This relation has to be valid for AB accompanying sound propagation.

Shock waves that are formed in gases or plasma by fast moving objects are characterized by rapid variation of the gas density and pressure at the front of the shock wave. It means that a shock wave has to be a source of AB as well.

Creations of cracks and splitting of material under mechanical tension becomes a source of radiation. It was demonstrated long ago that in the process of splitting of mica high-energy $\gamma$-rays are emitted with energies up to 3 MeV with a yield by about ten orders of magnitude higher than expected according to thermal equilibrium distribution (Perel'man and Khatiashvili, 1982).

When bodies are mechanically destroyed, on the sides of a crack oppositely charged layers are created. The relative motion of these layers generate radiation and can accelerate electrons that in their turn produce BrS in general and AB in particular. It seems that investigation of radiation from e.g. different metal construction can help to detect the tensions inside the construction and serve as a source of information on creation of cracks in material at early stages of destruction processes.

e. *Earth quakes*

Powerful radiation during earthquakes was registered for the first time in at the end of the seventies of the last century and since then it is studied as an important geophysical phenomenon. It is studied also with the aim to obtain a possibility to forecast earthquakes. Electromagnetic radiation is observed during all the seismic dislocations, in particular, during foreshocks and aftershocks of earthquakes. Concentration of electrical charges on the banks of the cracks and consequent recombination of these charges could only partially explain these phenomena. A theoretical approach to the generation of electromagnetic radiation that accompanies the process of mechanical destruction was proposed and applied to some simple systems relatively long ago by Khatiashvili and Perel'man (1989). It is worthwhile to note that formation and (or) oscillation of dislocations (see Perel'man and Khatiashvili, 1982) that are inevitably accompanied by AB play an important role in the earthquake radiation.

Destruction of materials very often forms double layers that are oppositely charged. They are similar to *electron-hole* pairs created under the action of a projectile in atomic collisions. Relative movement of such layers generates radiation. In rocks they the double layers are covered with water jackets that affects generation of radiation. Note, that double layers in living cells transforms e.g. acoustic into electro - magnetic waves (see Perel'man and Khatiashvili, 1983).

f. *Condensation of water and phase transitions*

Formation of liquids from gases and solids from liquids is accompanied by considerable modification of their electronic structure. These processes are inevitably accompanied by inducing of multipole, mainly dipole moments and therefore by creation of AB. However, the real role of AB in these processes is unclear. Having in mind, however, the big scales of such processes as e.g. condensation in nature, the investigation of the AB role in such processes is not only interesting, but of great importance. Indeed, the amount of condensing water on the Earth is about $10^{20}$ g per year. The radiation of latent heat of steam condensation is thus very big and an essential part of it goes to radiation.

g. *Radiation in the process of friction*

Microscopically, the friction of two bodies in contact that move relative to each other is accompanied by mechanical motion of irregularities of contacting surfaces. The motion of these irregularities lead to their polarization and as a result generates AB. A simplified version of such process, which can serve as a model of it, is radiation of an "atom" that moves parallel to a regular potential structure that is depicted in Fig. 6:

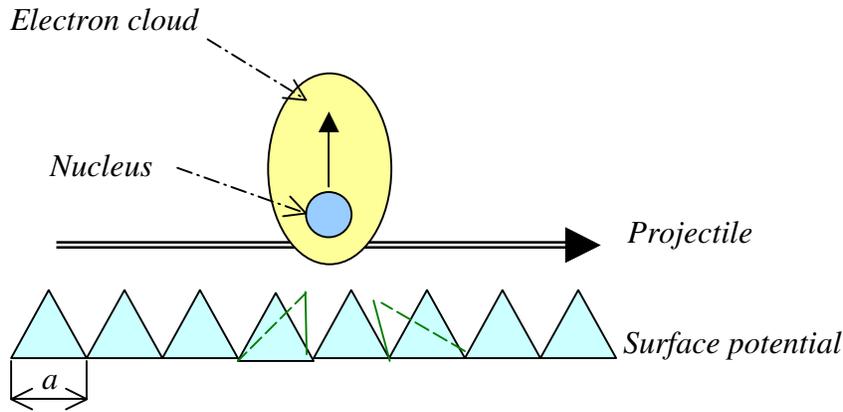

Fig. 6 Radiation of a charge near a saw-like surface potential

The radiation can be generated due to motion of a charge parallel to a "saw"-shape periodic field that was called "Smith-Purcell" radiation and discovered about half a century ago by Smith and Purcell (1953)). However it is possible that potential "teethes" can be deformed, as is shown in the figure, and becomes a source of some kind of AB.

If the projectile has internal structure, it can be dipolar polarized by the "saw" potential depicted in Fig. 6 and emit AB. It seems that the spectrum of it has to have a resonance at $\omega = v_p / a$, where $a$ determines the surface potential periodicity.

h. *Nuclear decay*

Nuclear decay processes are also accompanied by polarization radiation. Indeed, in a decay process a nuclear particle moves via the electron shells causing their deformation and as a consequence AB emission.

To be concrete, let us start with $\beta$- decay. Here one has several sources of radiation. At first, the electric charge of the nucleus is instantly altered due to transformation of e.g. one of nuclear neutrons into a proton: $n \to p + e + \bar{\nu}$. Instant variation of charge deforms or effectively polarizes the electron shells. However this source of BrS is very weak since the decay changes the field almost entirely spherically – symmetric, so that dipole excitations of the electrons are suppressed. However, instant change of the central field leads to so-called shake–off of atomic electrons that leave the atom relatively slow. BrS in general and AB in particular accompany this movement of electrons. The emission of a $\beta$- electron leads to recoil of the nucleus. That generates BrS in a way similar to that in neutron-atom collisions (see Section 7). The last source of BrS is the $\beta$-electron itself that produces both OB and AB while moving from the nucleus through the electron shells.

The $\alpha$ – decay as a source of radiation has some essential specific features as compared to the $\beta$ – decay. The $\alpha$-particle is by five orders of magnitude heavier than the $\beta$-electron. So, the recoil is considerably stronger, the instant change of the nuclear charge and corresponding shake-off probability is bigger. Due to big mass the only radiation generated by the $\alpha$-particle itself is AB.

Nuclear fission has to be accompanied by separation of electrons that stick to the fission products and become released. This separation generates radiation, a part of which can be considered as AB.

It seems that investigating the radiation from atomic electrons that accompanies the nuclear decay and fission can give valuable information about the corresponding nuclear processes.

Let us emphasize that in this subsection we have discussed the radiation that comes from deformation or polarization of electron shells of an atom (AB), the nucleus of which decays. Its typical $\omega$ is of the order of electron binding energies. However, radiation comes also from pure nuclear processes that have almost nothing to do with electron shells of the corresponding atom. The nuclear radiation that accompanies nuclear decay or fission is extensively investigated [see e.g. Tkalya, 1999, Maydanyuk and Olkhovsky, 2003.

i. *Laser ionization*

The action of an intense laser beam leads to multiple ionization of the illuminated atoms. For It appeared that for high enough intensities multiple ionization of atoms have relatively high probabilities. This was explained as a result of inelastic scattering of one initially ionized electron by its parent atom (see Kuchiev, 1999 and references therein). Oscillating in the laser field, this electron acquires a big *pondermotive energy* $E_{pon}$, $E_{pon} \approx I_L / \omega_L^2$, where $I_L$ and $\omega_L$ are the laser beam's intensity and frequency. The energy $E_L$ is much bigger than the ionization potential of the parent atom already for not too low $\omega_L$, $\omega_L = 1 eV$ and $I_L \geq 1 at.un.$ (1 *at. un.* of intensity is $10^{16} Watts/cm^2$). Bouncing in the laser field, this electron hits the parent atom, ionizing it or emitting BrS radiation. This process repeats times and again since the electron regains the lost energy from the laser beam. From the point of view of this paper it is essential that in each of these collisions AB can contribute essentially to the total BrS intensity.

## 11. "Exotic" processes

In this Section we will discuss generation of BrS in collisions that are difficult to study experimentally but are of interest to clarify the role and features of AB, or PR. We will also mention the possibility that other than photons particles can be created by a mechanism like AB.

Let us start with positron on atom, $e^+ + A$ scattering. According to (7), in this case the relative sign of OB and AB is opposite to that in $e^- + A$ case. Since a pair $e^+$ and $e^-$ has a dipole moment, even for fast $e^+ + A$ collisions and high $\omega$ there is no "atomic striptease". For the positron amplitude one has $F_{\vec{p}\vec{q}}^{(+)}(\omega) \sim [Z - 2Q(q)]$ instead of $F_{\vec{p}\vec{q}}^{(-)}(\omega) \sim Z$. For small $q$ $e^+$ and $e^-$ amplitudes differ only by their sign, since for a neutral atom $Q(0) = Z$. In slow $e^+ + A$ collisions the formation of real or virtual *Positronium Ps* affects the BrS cross-section essentially magnifying AM role (see Astapenko et al., 1985 and Amusia et al., 1992 and references there in).

Of prominent theoretical interest is such an exotic process as $e^{\pm} + Ps$ scattering. The distributed charge of *Ps* [see (4)] is equal to zero. Therefore all BrS in this collision comes from AB of *Ps* and from its excitation or disintegration. It is of interest to note that contrary to the fast $e^{\pm} + H$ collisions in $e^{\pm} + Ps$ the inelastic AB contributes at least not less than the elastic.

Since the distributed charge of *Ps* is zero, the BrS generated in $Ps + A$ collisions is according to (19) determined by polarization of *Ps* caused by the distributed charge of the atom $A$. As to radiation in a gas mixture consisting of *Ps* atoms in their ground and exited quadrupole states $Ps^{(2)*}$ (the dipole excitation is a short living), it is determined by the product $\alpha_d^{Ps}(\omega) Q_{Ps}^{(2)}(q)$, where $Q_{Ps}^{(2)}(q)$ is the form-factor of $Ps^{(2)*}$.

Of theoretical interest is also BrS generation in $e^{\pm} + H_{\mu}$ collisions. So-called muonic atom $H_{\mu}$ consists of a $\mu^-$-meson with mass $m_{\mu} \approx 210$ and a proton. The radius $a_{\mu}$ of $H_{\mu}$ is equal to $a_{\mu} = 1/m_{\mu}$. However even the low energy $e^{\pm} + H_{\mu}$ cross section is not of the order of $a_{\mu}^2$, but four orders of magnitude smaller, $a_{\mu}^2 / m_{\mu}^2$. At first glance since the incoming electron (positron) is a "light" particle in this collision, its OB has to dominate. But because $H_\mu$ is neutral and the elastic $e^\pm + H_\mu$ scattering cross section is very small, in fact at any $\omega$ except $\omega \ll m_\mu$ AB dominates (see Amusia and Soloviev, 1985; Amusia et al., 1992).

Another exotic object is the $\nu + H$ scattering (see Amusia et al., 1992 and references therein). Since neutrino has no electric charge, OB can come only due to its magnetic moment. If not zero, it mast be however very small. So, neutrino radiates only due to the AB mechanism, by inducing dipole moment in the target hydrogen atom because $\nu e$ and $\nu p$ interactions, or corresponding neutral currents, are different. The neutrino interactions are of short range and therefore, contrary to the $e + H$ case, big exchange of momentum $q$ are essential in AB generation. One has also to take into account an important contribution of $\nu + H$ collision with spin-flip of the atomic electron. It appeared that detection of AB resonant lines that come from excitation of the targets discrete levels could serve to determine the neutrino flux direction, when using polarized atomic hydrogen as a target. Note that at resonance frequencies the AB intensity is strongly enhanced, as compared to the non-resonant region. This enhancement reaches six orders of magnitude for incoming neutrino with the energy of 100 at. un.

Recent experimental production of anti-hydrogen atom, $\bar{H}$, gives another exotic object from the point of view of radiation generating. Of course, investigation of $\bar{H}$ is of primarily interest as fundamental object that help to clarify the symmetries in Nature. But $\bar{H} + H$ collision and its comparison with $H + H$ are of interest also from the BrS point of view. For fast but non-relativistic collisions dipole AB is not generated in $H + H$, while it is generated in $\bar{H} + H$. Of course, OB for these heavy particles is negligible. It is interesting to note that for ultra-relativistic collisions the difference between $\bar{H} + H$ and $H + H$ as it can be seen from (20) disappears.

Basically the same mechanism as AB that compliments OB in forming the BrS spectrum can be of importance in producing not only photons but also other Bose-particles, both with and without rest mass. The simplest example is a *phonon*, i.e. a long-wave oscillations of the crystal structure of a solid body. The principal difference between photons and phonons is that a phonon can be created by a particle freely moving in the solid body, while a photon cannot be emitted by a freely moving particle. It means that OB for phonons, $OB_{phon}$, is enhanced. However, $AB_{phon}$ also exists. The indirect, via a mechanism, similar to AB, creation of *excitons* is also possible. At a much smaller distance and much higher energy scales we have the problem of $\pi$-meson production in nucleon-nucleus scattering. Here along with direct creation of $\pi$-mesons by incoming nucleon a process similar to AB, namely $\pi$-mesons production via virtual nucleon-vacancy excitations of the nucleus, must be taken into account.

**12. Some macroscopic manifestations**

In this section we will discuss some macroscopic manifestations of polarization radiation. In several considered bellow examples it is essential that the target is really a multi-atomic object.

a. *Cooling of gaseous objects*

Consider two volumes of gas at temperature $T_0$ in boxes with outer transparent walls. We assume that $T_0$ is much higher than the temperature outside, $T_{out}$. The common wall has a closed door. The considered system is shown in Fig. 7a.

a) *Door is closed*

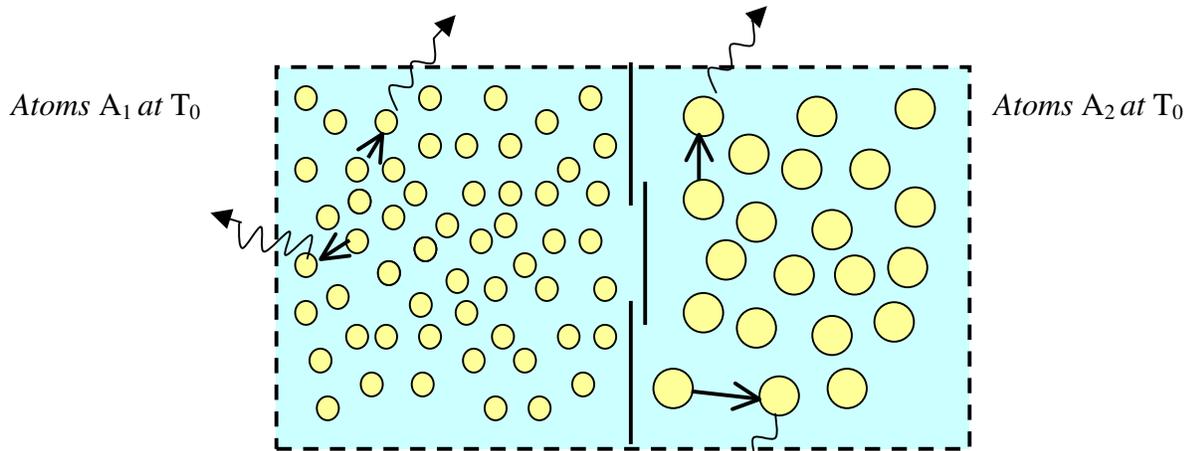

b) *Door is open*

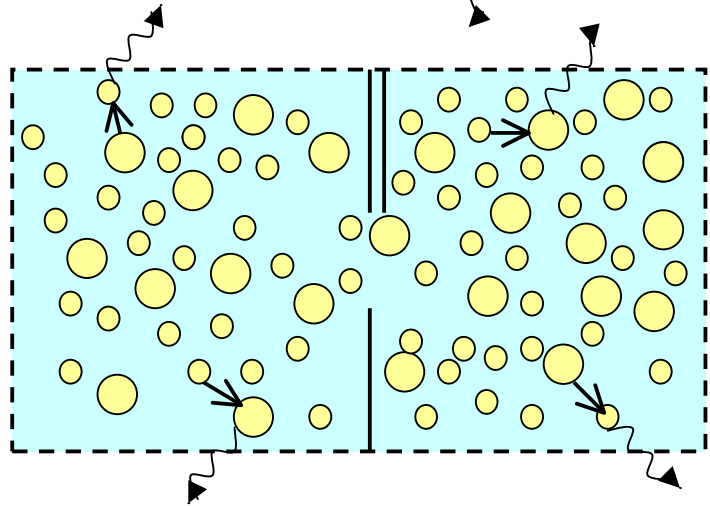

Fig. 7. Radiative cooling of a gas volume.
  a) No dipole radiation. b) Dipole radiation exists

To be concrete, assume that the left volume is filled with $A_1$ and the right with $A_2$ gases. When the temperature $T_0$ is low enough, in collisions the atoms cannot be ionized. So, the only mechanism of cooling, apart of transferring the heat of gas to the walls (which for big volumes is not important), is via AB emission. However, for a gas that consists of identical atoms, AB is according to (23) suppressed by a factor $(v/c)^2 \sim kT_0/c^2 \approx 10^{-6 \div 7}$, where $k$ is the Boltzmann constant. After opening the door (see Fig. 7b), gases $A_1$ and $A_2$ mix due to diffusion. AB then becomes intensified by a factor of $10^{+6 \div 7}$ for pairs of different atoms thus dramatically increasing the rate of radiative cooling. Such an experiment would serve as an impressive manifestation of AB effectiveness.

  b. *"Drug" currents of neutral particles*

It is usually assumed that light pressure, which causes directed motion of heavy ions or atoms under the action of a light beam is on the microscopic level based on photon elastic back scattering. However, as Amusia and Baltenkov (1986) have demonstrated, it exist another mechanism of light pressure that is efficient for heavy atoms or ions and is based on a process reversed to AB, namely on "Atomic" or polarization absorption, PA. Schematically this mechanism is depicted in the Fig. 8. An atom $A_1$ absorbs a photon and therefore dipole moment is induced in $A_1$. Due to collision with an atom $A_2$ not only energy $\omega$ but also momentum $\vec{k}, |k| = \omega/c$, is transferred, leading to directed motion of $A_2$. The directed motion of $A_2$ is preserved until its elastic collision with any target atom, of $A_1$ or $A_2$ type, happens. Note that for $A_1 = A_2$ PA is strongly suppressed due to the same reasons as AB [see (23)].

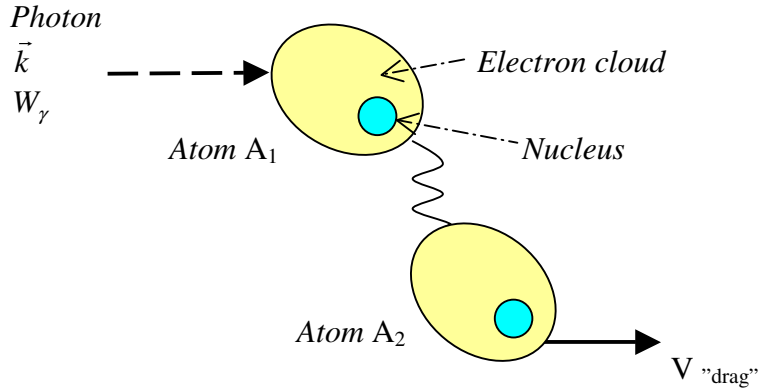

Fig. 8 Linear momentum transfer in Polarization Absorption

The flux of directionally moving atoms $f(\omega)$ is proportional to the product of photon flux $W_\gamma$, photon momentum $k$, and the PA cross-section $\sigma_{PA}(\omega)$ and inversely proportional to the elastic atom-atom scattering cross section $\sigma_{AA',el}(\omega)$:

$$f(\omega) \sim W_\gamma k \sigma_{PA}(\omega)/\sigma_{AA',el}(\omega). \quad (37)$$

Amusia and Baltenkov (1986) [see also Amusia and Baltenkov (2000)] obtained using (19) for a mixture of atoms with the polarizability $\alpha_d(\omega)$ and ions with the charge $Z$ that are at temperature equivalent to energy $E = v^2/2M$:

$$\sigma_{PA}(\omega) \sim \frac{\omega}{c}\left[\frac{Z\alpha_d(\omega)}{v}\right]^2 \ln\left(\frac{v}{\omega\varsigma_0}\right), \quad (38)$$

$$\varsigma_0 = [Z\alpha_d(0)/2E]^{1/4}.$$

The PA light pressure can be estimated using (37) and (38) as

$$P_{PA} \sim \sigma_{PA}(\omega) v n, \quad (39)$$

where $n$ is the target density. This pressure has to be compared with ordinary light pressure $P_O$, that can be estimated depending for structureless charges as $P_O \sim (8\pi/3c^4)(Z^2/M)^2$ or for neutral atoms $P_O \sim (8\pi/3c^4)\omega^4\alpha_d^2(\omega)$. Since $P_{PA}$ increases with density $n$ growth, it can become bigger than $P_O$. Let us estimate the density $n_{PA}$, starting from which $P_{AB} > P_O$. For $\omega = 0.1 eV$ and gas temperature $E = 5eV$ one obtains $n_{PA} \approx 10^{12} cm^{-3}$ that is the density of the e.g. upper atmosphere, comet tails etc.

c. *Cherenkov radiation of neutral particles*

It is well known that a particle that moves in a substance with a speed $v$ higher than the speed of light in this substance $c_{sub}$, $v > c_{sub}$ emits Cherenkov radiation. Amusia and Soloviev (1986) have demonstrated that a neutral atom also at $v > c_{sub}$ emits Cherenkov radiation, although of much lower intensity. It radiates since it has a form-factor and a quadrupole moment due to Lorenz contraction of the charge distribution. The incoming particle's energy loss per unit path is given by the following expression

$$\frac{dF^{(0)}}{d\omega} = \frac{Q_{(L)}^2 \omega^5}{16c^2v^4}\left(1-\frac{1}{\beta^2 n^2(\omega)}\right)\left(1-\frac{1}{n^2(\omega)}\right)^2,$$

$$Q_{(L)} = -2\beta^2 \left\langle 0\left|\sum_{j=1}^{N} z_j^2\right|0\right\rangle, \beta = v/c.$$

(40)

Here $Q_{(L)}$ is the momentum that appears due to Lorenz contraction in the laboratory frame and presence of non-zero form-factor of the spherically symmetric projectile in its rest frame that moves with the velocity $v \approx c$; $n(\omega)$ is the substance's refractive index, $\beta = v/c$. The value $dF^{(0)}/d\omega$ is, unfortunately, by $(kr_{at})^4 = (\omega r_{at}/c)^4 \sim 10^{-9}$ times smaller than for a charged particle, $dF^{(Z)}/d\omega$.

A natural question is whether an atom moving fast enough in a substance can remain to be neither excited nor ionized (see Ginzburg, 1981). It this connection it is essential to note that for $\gamma \equiv 1/\sqrt{1-\beta^2} \gg 1$ the projectile's ionization is suppressed since it requires for this very big energy $I_{(L)} = \gamma I_{proj}$, where $I_{proj}$ is the ionization /excitation potential of the atom in its rest frame.

### 13. Qualitative schemes of some other alternative mechanisms

We will describe here some classical mechanisms of generating of electromagnetic radiation, in which turning into motion or deforming of the target by a projectile creates the radiation. Fig. 9 depicts a dipole that is turned into rotation by an incoming charge. Note that the projectile can also emit radiation but we call attention to the fact that the rotating dipole, e.g. a dipole molecule, can be a source of radiation.

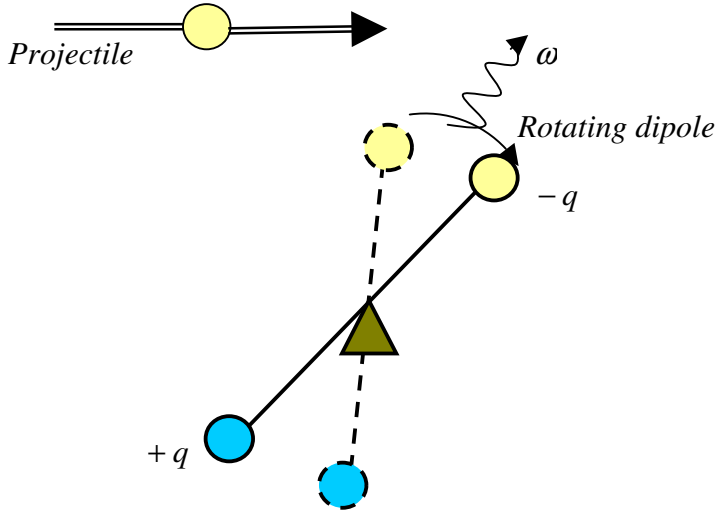

Fig. 9. A dipole turned into rotation by a projectile

This mechanism can be useful in modeling the radiation of dipolar molecules or highly excited atoms.

Fig. 10 depicts a more complex device, which consists of two dipoles. Without external action this system is not only neutral, but also has no dipole moment. But an incoming charge can deform the construction inducing a dipole moment. The latter will change in magnitude and direction as a function of the distance between the projectile and the target device. The space and time variation of the induced dipole moment of the target device led to emission of photons.

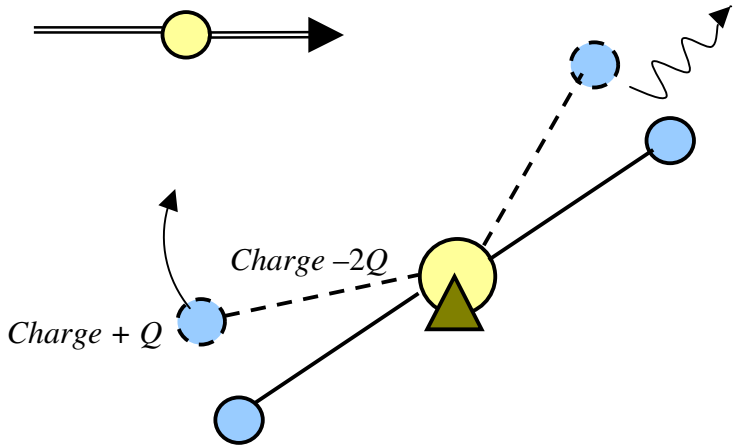

Fig. 10. Inducing of alternative dipole moment in a system of two dipoles.

Fig. 11 presents another classical mechanism: a magnetic arrow is turned into rotation by a moving magnet.

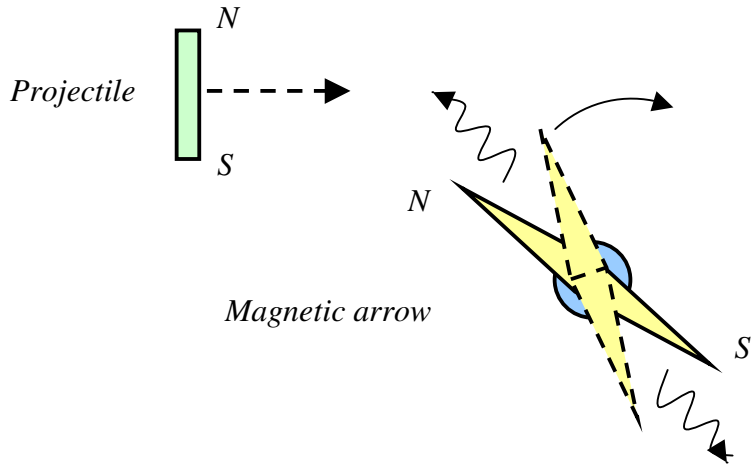

Fig. 11. A magnetic arrow, that rotates due to interaction with a moving magnet.

Of course, the mechanism Fig. 11 can be more complex and includes an option of the target magnet to be stretched under the action of the projectile. As a result the magnetic moment cannot only alter its direction but also the magnitude, thus in each case and in their combination leading to emission of radiation. Note that the magnetic arrow and static magnet can be arranged using closed and electrical currents.

In all these examples the photon energy is taken from the projectile's kinetic energy, but the photon is emitted by the dipole moment induced in the target, as it should be in AB. The microscopic and hence quantum-mechanical picture of the devices presented in Fig. 9-11 is not yet developed.

In principle, some other schemes to produce electromagnetic radiation could be suggested, namely those, in which the target would be not only deformed, but destructed. Such destruction would be inevitably accompanied by emission of radiation.

## 14. Conclusions

We have discussed here the foundations of Atomic Bremsstrahlung (AB) as a source of mainly continuous spectrum radiation and a really universal phenomenon that is essential from nuclei and nucleons (and their constituents) to atoms, molecules, clusters, fullerenes and macroscopic objects. We presented results of some calculations that show important and in some cases decisive role played by AB. We discussed, inevitably briefly, a whole variety of natural phenomena, where AB could be of importance. We presented non-trivial results for exotic objects that are far from being a possible direct object of experimental investigation but are of essential theoretical interest. We also discussed some new mechanisms, pure classical, where the target that is turned into motion in collision process, not the projectile, becomes a source of radiation.

Obviously, in this essay, which is not precisely what a review paper should be, far from all possibilities are discussed or even mentioned. At the beginning of the research in this field and even now, while thinking about AB, I imagine collisions of different objects, from quarks to Galaxies, where not only photons but other objects, e.g. gluons and gravitational waves are generated. Such fantasies are really inspiring and encouraging to see the World in its unity.

It is a lot to be done in this domain. What bothers however is that in spite of impressive activity from the side of theory there are quite a few experimental confirmations of AB or PB important role. The experimental verification of numerous theoretical predictions in this field would be very important and encouraging for the theorists. It would also help a lot in establishing AB as a specific and different from OB mechanism of radiation. I do hope that publication of a special issue of the Radiation Physics and Chemistry will stimulate the research activity in this domain.

**Acknowledgement**

I am grateful to the Binational Science Foundation, grant 200264, to the Israeli Science Foundation, grant 174/03 and to the Hebrew University Intramural fund for financial support of this research.